\documentclass[aps,twocolumn,prx]{revtex4-2}

\usepackage{xcolor,hyperref}
\hypersetup{
   colorlinks,
   linkcolor={blue!50!black},
   citecolor={blue!50!black},
   urlcolor={blue!80!black}
}

\usepackage{amsmath,bm,epsfig,color}
\usepackage{soul}
\usepackage{hyperref}
\usepackage{footmisc}
\usepackage{wrapfig}
\usepackage[normalem]{ulem}

\usepackage{algorithm}
\usepackage{algpseudocode}

\DeclareMathAlphabet{\mathitbf}{OML}{cmm}{b}{it}

\usepackage[latin1]{inputenc}
\newcommand{\beq}{\begin{equation}}
\newcommand{\eeq}{\end{equation}}
\newcommand{\bea}{\begin{eqnarray}}
\newcommand{\eea}{\end{eqnarray}}
\newcommand{\pa}{\partial}

\renewcommand{\=}{\!=\!}

\newcommand{\dbar}{{\,\mathchar'26\mkern-12mu d}}
\renewcommand{\th}{^{\mbox{\tiny th}}}
\newcommand{\dv}{\mathitbf d}
\newcommand{\xv}{\mathitbf x}
\newcommand{\nv}{\mathitbf n}

\newcommand{\uv}{\mathitbf u}
\newcommand{\rv}{\mathitbf r}

\newcommand{\psiv}{\bm{\psi}}

\newcommand{\calBold}[1]{\mbox{\boldmath${\cal #1}$}}

\begin{document}

\title{Anomalous linear elasticity of disordered networks}
\author{Edan Lerner$^{1}$}
\email{e.lerner@uva.nl}
\author{Eran Bouchbinder$^{2}$}
\email{eran.bouchbinder@weizmann.ac.il}
\affiliation{$^{1}$Institute for Theoretical Physics, University of Amsterdam, Science Park 904, 1098 XH Amsterdam, the Netherland\\
$^{2}$Chemical and Biological Physics Department, Weizmann Institute of Science, Rehovot 7610001, Israel}

\begin{abstract}
Continuum elasticity is a powerful tool applicable in a broad range of physical systems and phenomena. Yet, understanding how and on what scales material disorder may lead to the breakdown of continuum elasticity is not fully understood. We show, based on recent theoretical developments and extensive numerical computations, that disordered elastic networks near a critical rigidity transition, such as strain-stiffened fibrous biopolymer networks that are abundant in living systems, reveal an anomalous long-range linear elastic response below a correlation length. This emergent anomalous elasticity, which is non-affine in nature, is shown to feature a qualitatively different multipole expansion structure compared to ordinary continuum elasticity, and a slower spatial decay of perturbations. The potential degree of universality of these results, their implications (e.g.~for cell-cell communication through biological extracellular matrices) and open questions are briefly discussed.
\end{abstract}

\maketitle


Continuum elasticity is a powerful tool that describes a huge range of phenomena in diverse physical systems, over sufficiently large lengthscales~\cite{landau_lifshitz_elasticity}. On sufficiently small lengthscales, where the effect of material structure and disorder cannot be coarse-grained, continuum elasticity theory is expected to break down. The crossover between these two regimes is characterized by a correlation length $\xi$. Indeed, recent work has identified $\xi$ for glassy systems, particle packings and elastic networks~\cite{fred_pre_2005,phonon_gap_soft_matter_2013,breakdown,gustavo_strain_stiffening_pre2014,atsushi_core_size_pre,sticky_spheres1_karina_pre2021,pinching_pnas}, and demonstrated the validity of continuum elasticity theory for lengthscales $r$ that satisfy $r\!\gg\!\xi$. In particular, the displacement response to a point force (monopole) perturbation decays as $1/r$ in three dimensions (3D) for $r\!\gg\!\xi$, while the displacement response to a force dipole --- which is proportional to the spatial gradient of the point force response~\cite{landau_lifshitz_elasticity} --- decays as $1/r^2$. Yet, it remains unclear whether a generic elastic response exists for $r\!\ll\!\xi$ and if so, what form it takes.

For systems undergoing a glass transition upon cooling a melt, $\xi$ has been shown to be of the order of 10 atomic distances and to vary mildly with glass formation history (e.g.~the cooling rate)~\cite{footnote}. Consequently, it is difficult to imagine that a generic elastic response emerges on such a narrow range of scale and one expects that disorder and fluctuations dominate the elastic response for $r\!<\!\xi$ in such systems. On the other hand, systems that undergo a critical rigidity transition --- such as fibrous biopolymer networks that are abundant in living systems (e.g.~collagen, fibrin and basement membrane) and that are known to undergo a dramatic stiffening transition when deformed to large enough strains~\cite{robbie_nature_physics_2016} --- feature a macroscopically large $\xi$ close to the rigidity transition~\cite{gustavo_strain_stiffening_pre2014,robbie_strain_stiffening_pre2018}. In such cases, the regime $r\!\ll\!\xi$ spans many orders of magnitude and might possibly accommodate a generic elastic response. In this brief report, we consider such disordered networks near their rigidity transition and study their elastic response for~$r\!\ll\!\xi$.


We show, based on recent theoretical developments and extensive numerical simulations, that the \emph{linear} elastic response of disordered networks follows an anomalous power-law for $r\!\ll\!\xi$, when $\xi$ is sufficiently large, and that the decay is slower compared to the continuum elastic response for $r\!\gg\!\xi$. Furthermore, we show that the anomalous elasticity for $r\!\ll\!\xi$ features a qualitatively different multipole expansion structure compared to continuum elasticity. These results may have significant implications for long-range mechanical interactions between distant cells through biological extracellular matrices~\cite{Janmey_2021_soft_matter}, inspire the design of heterogeneous structures with unusual properties~\cite{Acuna2022} and pose new basic questions, which are briefly discussed.

Before reviewing our results, we emphasize that many previous efforts to explain anomalous elastic responses in various biophysical contexts invoke intrinsically \emph{nonlinear} constitutive laws, see e.g.~\cite{sam_prl_2012,wang_2014,sam_pre_2015,yair_pre_2021,lenz_arXiv_2022}. Contrary to those studies, in this work we consider model systems in which an anomalous \emph{linear} response emerges, as also recently shown to occur in experiments on tensed fibrous hydrogels~\cite{shahar_2022}. We finally note that other approaches termed `anomalous elasticity' --- describing different mechanical phenomena in amorphous materials --- were recently put forward~\cite{itamar_screening_1,itamar_screening_2,itamar_screening_3}.\\

\noindent{\bf \large Main result}\\

\vspace{-0.3cm}

\noindent We consider isotropic disordered networks whose nodes are connected by relaxed Hookean springs and whose proximity to a rigidity (jamming-unjamming) transition is controlled by the connectivity $z$ (average springs per node), which is close to (yet larger than) the critical Maxwell threshold $z_{\rm c}\=2\dbar$ (where $\dbar$ is the spatial dimension). Such disordered elastic networks feature a shear modulus that scales linearly with $z-z_{\rm c}$~\cite{matthieu_thesis} and
\begin{equation}
\xi \sim \frac{1}{\sqrt{z-z_{\rm c}}} \ ,
\end{equation}
which diverges as $z\!\to\!z_{\rm c}$~\cite{silbert_length_pre_2005,liu_transverse_length_2013,breakdown,atsushi_core_size_pre}. Several additional growing lengthscales were previously identified near the unjamming transition; a comprehensive review of those can be found in~\cite{quasilocalized_states_of_self_stress}. Our goal is to understand the elastic response of disordered elastic networks for $r\!\ll\!\xi$.

The basic quantity we focus on is the displacement response ${\bm u}({\bm r})$ to a force dipole applied at the origin. The reason we consider the dipole response (and not, for example, the monopole response) is three-fold. First, our results are relevant for fibrous biopolymer networks, which constitute extracellular matrices to which cells adhere in physiological contexts. Adherent cells apply to their surrounding extracellular matrices contractile forces that are predominantly dipolar~\cite{Adherent_cells_2013}. Second, contact formation between solid particles during dense suspension flows, where the overdamped response is analogous to elastic response, generates a force dipole~\cite{gustavo_asm_collisions_pre2016}. Finally, in many cases low-energy excitations in disordered systems are of dipolar nature, e.g.~the universal nonphononic excitations in glasses~\cite{JCP_Perspective}.

Consider then the response function $C(r)\sim \langle{\bm u}({\bm r})\!\cdot\!{\bm u}({\bm r})\rangle$, where $\langle\cdot\rangle$ stands for an angular average, rendering $C(r)$ a function of the distance $r$ alone, and note that ${\bm u}({\bm r})$ is normalized such that $C(r)$ is dimensionless. Next, one can consider the integral $\int_0^{\xi} \!C(r)\, d^{\dbar}r$ in $\dbar$ spatial dimensions and ask about its scaling with $\xi$. If $C(r)\!\sim\!\exp(-r/\xi)$ for $r\!<\!\xi$, then one obtains $\int_0^{\xi} \!C(r)\, d^{\dbar}r\!\sim\!\int_0^{\xi} \!C(r)\, r^{\dbar-1} dr\!\sim\!\xi^\dbar$, i.e.~the naive scaling with the correlation length $\xi$. However, recent work~\cite{gustavo_asm_collisions_pre2016,new_variational_argument_epl_2016,atsushi_core_size_pre,mw_mean_field_description_pre2022} suggested that in fact
\begin{equation}
\label{eq:correlation_volume_length}
\int_0^{\xi} \!C(r)\, d^{\dbar}r \sim \xi^2 \ ,
\end{equation}
in any dimension $\dbar$.

The latter indicates the existence of long-range correlations, i.e.~that in fact $C(r)\!\sim\!r^{-2\beta}\exp(-r/\xi)$ for $r\!<\!\xi$, where the exponential $\exp(-r/\xi)$ represents the fact that continuum linear elasticity, which features a different power law, dominates the response for $r\!\gg\!\xi$. Consistency with Eq.~(\ref{eq:correlation_volume_length}) requires that $\beta\=(\dbar-2)/2$~\cite{gustavo_asm_collisions_pre2016}. Consequently, in 3D ($\dbar\=3$) we have for the dipolar displacement response
\begin{equation}
\label{eq:displacment}
{\bm u}({\bm r}) \sim \begin{cases}  \frac{\displaystyle 1}{\displaystyle \sqrt{r}} & \mbox{for}\quad r \ll \xi \\\\ \frac{\displaystyle 1}{\displaystyle r^{2}} & \mbox{for}\quad r \gg \xi \end{cases} \ ,
\end{equation}
where for $r\!\gg\!\xi$ we just have the continuum elastic dipole response. Note that the dipolar response is predicted in Eq.~(\ref{eq:displacment}) to decay significantly slower for $r\!\ll\!\xi$ than for $r\!\gg\!\xi$. The same argument leading to Eq.~(\ref{eq:displacment}) has been spelled out in the context of flowing dense suspensions~\cite{gustavo_asm_collisions_pre2016} (see the ``Discussion and Conclusions'' section therein, where the overdamped --- not the elastic --- response has been considered). A similar anomalous elastic response has been recently suggested using a Ginzburg-Landau description of the so-called overlap free-energy in a mean-field approach to glasses~\cite{mw_mean_field_description_pre2022}.
\begin{figure}[ht!]
\centering
\includegraphics[width=1.0\linewidth]{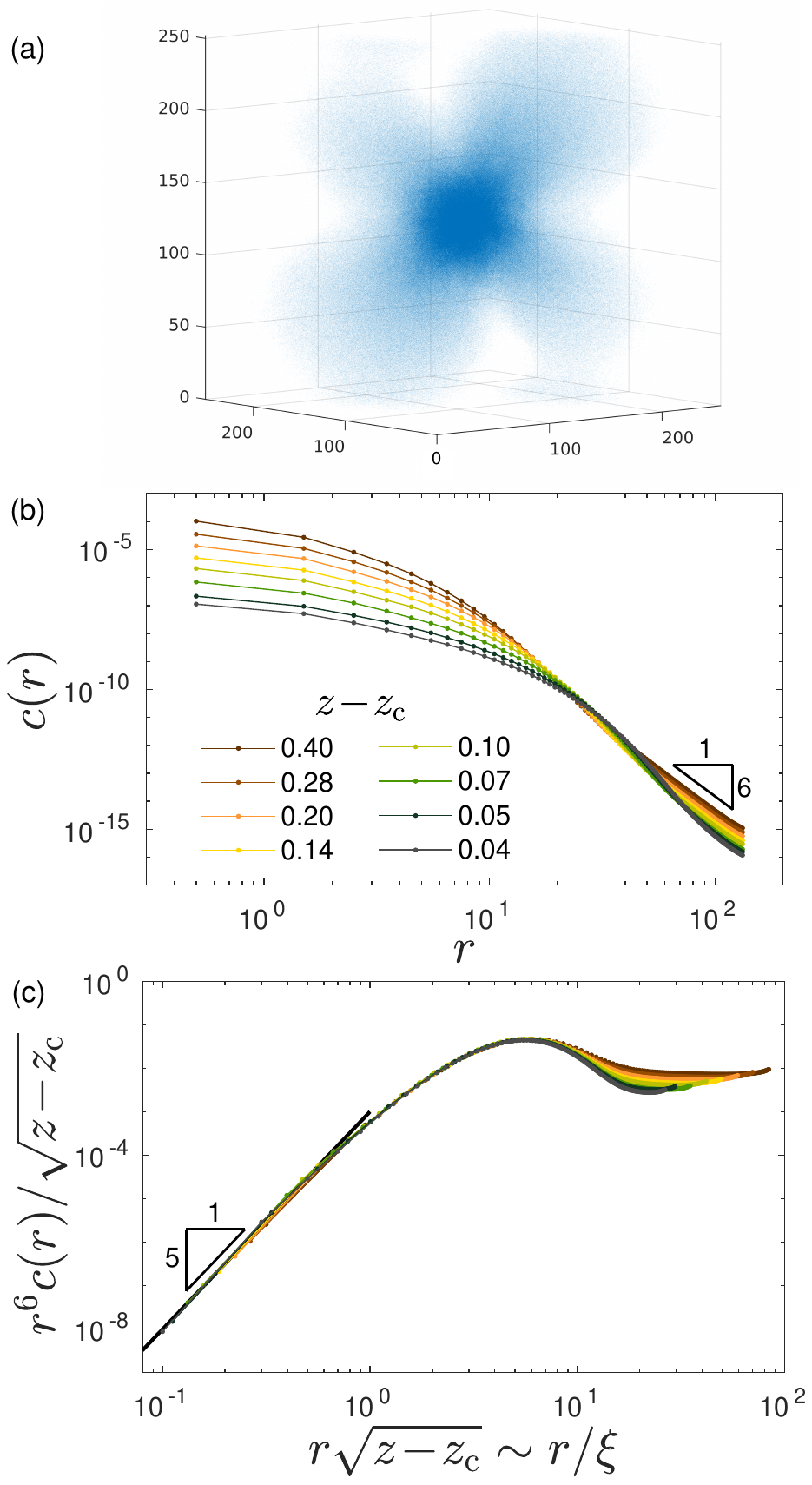}
\caption{\footnotesize (a) An example of the dipole displacement response $\uv(\rv)$, calculated in a spring network of $N\!=\!16$M nodes and connectivity $z\!=\!6.04$. The  inner, denser region corresponds to the anomalous elastic response, while the outer region corresponds to the continuum elastic response (which also exhibits the well-known quadropular symmetry~\cite{JCP_Perspective}, to be contrasted with the much more isotropic anomalous response). (b) Response functions $c(r)$ computed in networks of various connectivities $z\!-\!z_{\rm c}$, as indicated by the legend. For large $r$, $c(r)$ approaches the $r^{-6}$ scaling of continuum elasticity (cf.~Eq.~(\ref{eq:c(r)})). (c) The same data as in panel (b), but here plotted as $r^6 c(r)/\sqrt{z-z_{\rm c}}$ vs.~$r\sqrt{z-z_{\rm c}})^5\!\sim\!r/\xi$. The re-scaled representation reveals excellent collapse, which confirms that $(i)$ the correlation length follows $\xi\!\sim\!1/\sqrt{z\!-\!z_{\rm c}}$ and that $(ii)$ $c(r)\!\sim\!1/r$ for $r\!\ll\!\xi$, as hypothesized in Eq.~(\ref{eq:c(r)}).}
\label{fig:result}
\end{figure}
\vspace{-0.2cm}

To test the prediction in Eq.~(\ref{eq:displacment}), we set out to construct a response function that is slightly different than $C(r)$ (yet, see results for $C(r)$ in the {\color{blue}{Supplemental Materials}}). Our motivation for doing so is that we are not only interested in validating the form of ${\bm u}({\bm r})$ for $r\!\ll\!\xi$, but also in gaining insight into the structure of elasticity theory on these scales and the possible differences compared to continuum elasticity. The latter~features~a multipole expansion in which the n$\th$ multipole order involves a spatial gradient $\pa$ of the (n-1)$\th$ one. Consequently, we construct a response function $c(r)$ (see {\color{blue}{Supplemental Materials}} for details) that scales as $|\partial {\bm u}({\bm r})|^2$ for $r\!\gg\!\xi$, and explore whether the corresponding structure persists also for $r\!\ll\!\xi$. Moreover, $|\partial {\bm u}({\bm r})|^2$ is proportional to the elastic energy density of the dipolar response (to quadratic order), and hence is a directly relevant physical quantity.

The response for $r\!\ll\!\xi$ is expected to be strongly affected by disorder, and hence to be non-affine in nature~\cite{Ellenbroek_2009} and dominated by fluctuations~\cite{phonon_gap_soft_matter_2013}. Consequently, we expect the angular average of spatial derivatives of the displacement vector ${\bm u}({\bm r})$ to feature vectorial cancellations. As a result, one can \textit{hypothesize} that $c(r)$ for $r\!\ll\!\xi$ inherits its scaling from $|{\bm u}({\bm r})|^2$, not from $|\partial {\bm u}({\bm r})|^2$. If true, then $c(r)$ takes the form
\begin{equation}
\label{eq:c(r)}
c(r) \sim \begin{cases}  |{\bm u}({\bm r})|^2\!\sim\!\frac{\displaystyle 1}{\displaystyle r} & \mbox{for}\quad r \ll \xi \\\\ |\partial {\bm u}({\bm r})|^2\!\sim\!\frac{\displaystyle 1}{\displaystyle r^{6}} & \mbox{for}\quad r \gg \xi \end{cases} \ ,
\end{equation}
where $c(r)\!\sim\!|\partial {\bm u}({\bm r})|^2\!\sim\!r^{-6}$ for $r\!\gg\!\xi$ follows (by the construction of $c(r)$) from ${\bm u}({\bm r})\!\sim\!r^{-2}$ of Eq.~(\ref{eq:displacment}). Consequently, Eq.~(\ref{eq:c(r)}) suggests that the elastic response for $r\!\ll\!\xi$ is anomalous not just in its power-law decay, as predicted in Eq.~(\ref{eq:displacment}), but also as it might not follow the multipole expansion structure of continuum elasticity.

Quantitatively testing the prediction in Eq.~(\ref{eq:c(r)}) requires very large networks that span a broad range of lengthscales and feature a sufficiently large correlation length $\xi$, which is controlled by $z-z_{\rm c}$. To that aim, we generated disordered elastic networks of 16 million nodes each for various $z-z_{\rm c}$ values (see {\color{blue}{Supplemental Materials}}), and calculated their dipole response and subsequently $c(r)$ following its network-level definition (see {\color{blue}{Supplemental Materials}}, where we also show $C(r)$). A single dipole response is shown in Fig.~\ref{fig:result}a. We first aim at verifying the continuum elastic response in Eq.~(\ref{eq:c(r)}) for sufficiently large scales, i.e.~for $r\!\gg\!\xi$. In Fig.~\ref{fig:result}b, we plot $c(r)$ for several small values of $z-z_{\rm c}$ (as indicated in the legend). We find that indeed for sufficiently large $r$, all curves follow $c(r)\!\sim\!r^{-6}$ (and, as expected, more so the larger $z\!-\!z_{\rm c}$ is, corresponding to smaller $\xi$).

We then set out to test the prediction in Eq.~(\ref{eq:c(r)}) for $r\!\ll\!\xi$, which was our major goal. To that aim, we first note that assuming the spatial dependence of $c(r)$ in this regime (cf.~Eq.~(\ref{eq:c(r)})), one can also predict its $z\!-\!z_{\rm c}$ dependence. The result reads $c(r)\!\sim\!(z\!-\!z_{\rm c})^3/r$ for $r\!\ll\!\xi$ (see {\color{blue}{Supplemental Materials}}). The latter implies that $r^6 c(r)/\sqrt{z\!-\!z_{\rm c}}\!\sim\!(r\sqrt{z\!-\!z_{\rm c}})^5\!\sim\!(r/\xi)^5$ for $r\!\ll\!\xi$. This prediction is tested in Fig.~\ref{fig:result}c and is shown to be in excellent quantitative agreement with the direct numerical calculations.

\noindent{\bf \large Discussion}\\
\vspace{-0.4cm}

\noindent The results presented above, based on recent theoretical developments and extensive numerical simulations, provide strong evidence for the existence of anomalous elasticity in disordered networks close to their rigidity transition, below a correlation length $\xi$ that is expected to be macroscopically large. The elastic response for $r\!\ll\!\xi$ is anomalous, i.e.~different from ordinary continuum elasticity that is valid for $r\!\gg\!\xi$, in at least two major respects; first, the spatial decay of perturbations is \textit{slower} (i.e.~characterized by a smaller inverse power-law exponent) compared to continuum elasticity. Second, the elastic response for $r\!\ll\!\xi$ is highly non-affine and hence does not appear to follow the ordinary multipole expansion structure of continuum elasticity, where a higher order multipole response is obtained by spatial gradients of a lower order one. 

Recent experiments on tensed fibrin hydrogels~\cite{shahar_2022} show that the \emph{linear} response~\cite{footnote2} to a point force is anomalous, decaying as $1/r^\beta$ away from the perturbation --- with $\beta$ measurably smaller than 1 ---, instead of the $1/r$ decay expected from continuum linear elasticity. These results indicate that the mechanism generating anomalous elastic responses might not be the intrinsic nonlinearity of the constituent elements, as invoked in several previous works~\cite{sam_prl_2012,wang_2014,sam_pre_2015,yair_pre_2021,lenz_arXiv_2022}, but rather the physics discussed here. In this case, anomalous vibrational modes~\cite{matthieu_thesis} --- and not long wavelength (wave-like) modes --- dominate the response at distances $r\!<\!\xi$.

The anomalous elasticity discussed here may have significant implications for various systems such as flowing dense suspensions~\cite{gustavo_strain_stiffening_pre2014} (where the overdamped response is analogous to the elastic response) and strain-stiffened fibrous biopolymer networks that are abundant in living systems~\cite{robbie_nature_physics_2016,robbie_strain_stiffening_pre2018}, e.g.~as extracellular matrices to which cells adhere and apply contractile forces~\cite{Adherent_cells_2013}. In the latter context, our results may imply that for $r\!\ll\!\xi$ distant cells can mechanically communicate over significantly longer distances~\cite{Janmey_2021_soft_matter} (compared to continuum elasticity) using their active contractility, a capability that might be important in various physiological processes (e.g.~tissue development).


Our results also pose new questions and open the way for additional research directions. First, we focused here on the response to force dipoles, which is well motivated from the physical and biological perspectives, as discussed above. Yet, force monopoles (point forces) are of interest as well, both because they are of practical relevance and because of the unusual multipole structure of the anomalous elastic response we discussed. In particular, in view of our findings, one can speculate that the monolpole response is identical to the dipole response scaling-wise, which should be tested in future work. Second, our analysis considered the amplitude squared of the response, i.e.~$|{\bm u}({\bm r})|^2$. It would be interesting to understand whether the same scaling remains valid for the \emph{average} of the vectorial response itself, i.e.~${\bm u}({\bm r})$, which is obviously subjected to vectorial cancellations due to the non-affine nature of the response, see for example the results of~\cite{phonon_gap_soft_matter_2013}. Finally, the degree of universality of our results should be also tested in future work, e.g.~in different disordered materials. In particular, we speculate that the same anomalous decay observed here below $\xi$ also characterizes the universal spatial structure of soft quasilocalized excitations that generically emerge in structural glasses~\cite{JCP_Perspective}, and feature spatial structures that closely resemble responses to force dipoles~\cite{cge_paper}. 


\vspace{0.1cm}
\noindent\textit{\bf Acknowledgements.} We thank J.~Zylberg for developing and implementing the bond-dilution algorithm used in this work. E.L.~acknowledges support from the NWO (Vidi grant no.~680-47-554/3259). E.B.~acknowledges support from the Ben May Center for Chemical Theory and Computation and the Harold Perlman Family.
\vspace{-0.65cm}

\onecolumngrid
\begin{center}
	\textbf{\large Supplemental Materials for:\\``Anomalous elasticity of disordered networks''}
\end{center}

\setcounter{equation}{0}
\setcounter{figure}{0}
\setcounter{section}{0}
\setcounter{subsection}{0}
\setcounter{table}{0}
\setcounter{page}{1}
\makeatletter
\renewcommand{\theequation}{S\arabic{equation}}
\renewcommand{\thefigure}{S\arabic{figure}}
\renewcommand{\thesubsection}{S-\Roman{subsection}}
\renewcommand*{\thepage}{S\arabic{page}}
\twocolumngrid

\title{\textbf{\large Supplemental Materials for:\\``Anomalous elasticity of disordered networks''}}
\author{Edan Lerner}
\email{e.lerner@uva.nl}
\affiliation{Institute for Theoretical Physics, University of Amsterdam, Science Park 904, 1098 XH Amsterdam, the Netherlands}
\author{Eran Bouchbinder}
\email{eran.bouchbinder@weizmann.ac.il}
\affiliation{Chemical and Biological Physics Department, Weizmann Institute of Science, Rehovot 7610001, Israel}

\maketitle

\section{T\lowercase{he response function $c(r)$ and its scaling properties}}

We construct the response function $c(r)$ as follows:
\begin{enumerate}
    \item We define the dimensionless unit force dipole $\dv^{(ij)}$ applied on the pair of nodes $i,j$ as
    \begin{equation}
    \label{eq:d_def}
        \dv^{(ij)}_k = \frac{\partial r_{ij}}{\partial\xv_k} = (\delta_{jk}-\delta_{ik})\,\hat{\nv}_{ij}\,,
    \end{equation}
    where the $k\th$ node coordinates are denoted by $\xv_k$, $r_{ij}\!\equiv\!|\xv_{ij}|$ is the (scalar) pairwise distance between nodes $i$ and $j$, $\xv_{ij}\!\equiv\!\xv_j\!-\!\xv_i$ is the vector difference, and $\hat{\nv}_{ij}\!\equiv\!\xv_{ij}/r_{ij}$ is the unit vector pointing from node $i$ to node $j$.

    We note that the contraction of the dipole $\dv$ --- as defined in Eq.~(\ref{eq:d_def}) above --- with a field corresponds to taking the difference of the field across a bond, i.e.~to a discrete (network-level) gradient in the bond direction.

    \item The response of the network to a unit dimensionless force dipole $\dv^{(ij)}$ acting on the $(ij)$ spring is calculated through
    \begin{equation}
        \uv^{(ij)} = \calBold{M}^{-1}\cdot\dv^{(ij)}\,,
    \end{equation}
    where $\calBold{ M}\!\equiv\!\frac{\partial^2U}{\partial\xv\partial\xv}$ is the Hessian matrix. An example of such a dipolar response is presented in Fig.~1a in the manuscript. We then define the normalized response as $\hat{\uv}^{(ij)}\!\equiv\!\uv^{(ij)}/\sqrt{\uv^{(ij)}\cdot\uv^{(ij)}}$ and subsequently use it.
    \item We are next interested in the \textit{extension/compression} of each spring connecting nodes $m,n$, associated with the normalized response $\hat{\uv}^{(ij)}$ to a dimensionless unit force dipole $\dv^{(ij)}$ applied on the pair of nodes $i,j$.
    This quantity is given by
    \begin{equation}
    \label{eq:A_ij}
        A_{(ij),(nm)} = \hat{\uv}^{(ij)}\cdot\dv^{(nm)} = \frac{\dv^{(ij)}\cdot\calBold{M}^{-1}\cdot\dv^{(nm)}}{\sqrt{\dv^{(ij)}\cdot\calBold{M}^{-2}\cdot\dv^{(ij)}}}\ .
    \end{equation}
    Recall that a contraction of a vectorial field with $\dv^{(nm)}$ corresponds to the difference of the vectorial field across the nodes $m,n$, projected on the direction of the $(mn)$ bond.
    \item We are then interested in the average dimensionless energy density associated with the bond extension/compression of Eq.~(\ref{eq:A_ij}) at a distance $r$ from the position where the dimensionless unit force dipole $\dv^{(ij)}$ is applied. This quantity, denoted by $c_{ij}(r)$, is given (to quadratic order) by
    \begin{equation}
        c_{ij}(r) = \langle  A_{(ij),(nm)}^2\rangle_{_{r_{ij,nm}=r}}\,,
    \end{equation}
    which is an angular average of $A^2_{(ij),(nm)}$ over all $(mn)$ bonds at a distance $r$ from the $(ij)$ bond.
    \item Finally, the response function $c(r)$ reported in Fig.~1b-c in the manuscript is given by an average of the individual functions $c_{ij}(r)$ over a large set of random edges $i,j$.
\end{enumerate}

In Fig.~1c in the main text, we presented a scaling collapse of the products $r^6c(r)$ measured in 3D relaxed Hookean spring networks. This is achieved by rescaling the abscissa by $1/\sqrt{z\!-\!z_{\rm c}}$ (since the correlation length $\xi\!\sim\!1/\sqrt{z\!-\!z_{\rm c}}$) and the ordinate by $\sqrt{z\!-\!z_{\rm c}}$. The latter rescaling is motivated as follows.

First, we write the sum-of-squares of the dipole displacement response function $\uv$ as (omiting the $(ij)$ superscript for the ease of notation)
\begin{equation}
\label{eq:sum_of_squares_a}
    \uv\cdot\uv = \dv\cdot\calBold{M}^{-2}\cdot\dv =  \sum_\ell \frac{(\psiv_\ell\cdot\dv)^2}{\omega_\ell^4}\,,
\end{equation}
where $\psiv_\ell$ are the eigenfunctions of the Hessian $\calBold{M}$, $\omega_\ell^2$ are the eigenvalues associated with the eigenfunctions $\psiv_\ell$ and $\dv$ is a dimensionless unit force dipole. It is known that in relaxed Hookean spring networks with $z\!\to\!z_{\rm c}$, one has $\psiv_\ell\!\cdot\!\dv\!\sim\!\omega_\ell$~\cite{silbert_pre_2016}. The sum above can thus be approximated by an integral over the vibrational density of states as~\cite{breakdown}
\begin{equation}
\label{eq:sum_of_squares}
    \sum_\ell \frac{(\psiv_\ell\cdot\dv)^2}{\omega_\ell^4}\sim \int_{\omega_\star}\frac{\omega^2\,{\cal D}(\omega)}{\omega^4} \sim \frac{1}{\omega}\bigg|_{\omega_\star}\sim \frac{1}{z-z_{\rm c}}\,,
\end{equation}
where $\omega_\star$ is a characteristic frequency in the unjamming of relaxed Hookean spring networks, whose scaling $\omega_\star\!\sim\!z\!-\!z_{\rm c}$ is well established~\cite{new_variational_argument_epl_2016_SM}.

Combining the results in Eqs.~\eqref{eq:sum_of_squares_a}-\eqref{eq:sum_of_squares}, i.e.~$\uv\cdot\uv\!\sim\!(z-z_{\rm c})^{-1}$, with the spatial scaling of $\uv({\bm r})$ discussed in the manuscript, $\uv({\bm r})\!\sim\!r^{-(\dbar-2)/2}$, we obtain for the \textit{normalized} response $\hat{\uv}(\rv)$ the following prediction
\begin{equation}
|\hat{\uv}(\rv)| \sim \sqrt{z-z_{\rm c}}\,|\uv(\rv)|\sim \frac{\sqrt{z-z_{\rm c}}}{r^{(\dbar-2)/2}}\,.
\end{equation}

To proceed, we consider the energy $e$ associated with a unit dimensionless dipole response $\uv$ in a relaxed spring network, summed over all springs \emph{apart} from the perturbed spring. It was shown that~\cite{quasilocalized_sss_epje2018_SM}
\begin{equation}
    e \sim \int_{a_0}^\xi (\uv\cdot\dv)^2 \,r^{\dbar-1}dr \sim z-z_{\rm c}\,,
\end{equation}
where $a_0$ is the bond length. If we then invoke the \textit{hypothesis} that $\uv\!\cdot\!\dv\!\sim\!1/r^{(\dbar-2)/2}$, spelled out in the manuscript, and account for coordination-dependence of $\uv\!\cdot\!\dv$ by setting $(\uv\!\cdot\!\dv)^2\!\equiv f(z)/r^{\dbar-2}$, we obtain
\begin{equation}
    e \sim f(z)\int_1^\xi r^{-(\dbar-2)} \,r^{\dbar-1}dr \sim f(z)\,\xi^2 \sim z-z_{\rm c}\,.
\end{equation}

The last relation, together with $\xi\!\sim\!1/\sqrt{z\!-\!z_{\rm c}}$, implies that $f(z)\!\sim\!(z\!-\!z_{\rm c})^2$. Consequently, $A_{(ij),(nm)}$ defined in Eq.~\eqref{eq:A_ij}, which scales as $\uv\!\cdot\!\dv$, satisfies
\begin{equation}
    A_{(ij),(nm)}  \!\sim \frac{\sqrt{z-z_{\rm c}}\sqrt{f(z)}}{r^{(\dbar-2)/2}}\sim \frac{(z-z_{\rm c})^{3/2}}{r^{(\dbar-2)/2}}\,.
\end{equation}
Finally, since $c(r)$ scales as $(A_{(ij),(nm)})^2$, we obtain
\begin{equation}
    c(r) \sim (z-z_{\rm c})^3\,r^{-(\dbar-2)}\,,
\end{equation}
for $r\!<\!\xi$, which was used in the manuscript in 3D ($\dbar\=3$).

\section{T\lowercase{he response function} $C(r)$}

\begin{figure}[ht!]
 \includegraphics[width = 0.50\textwidth]{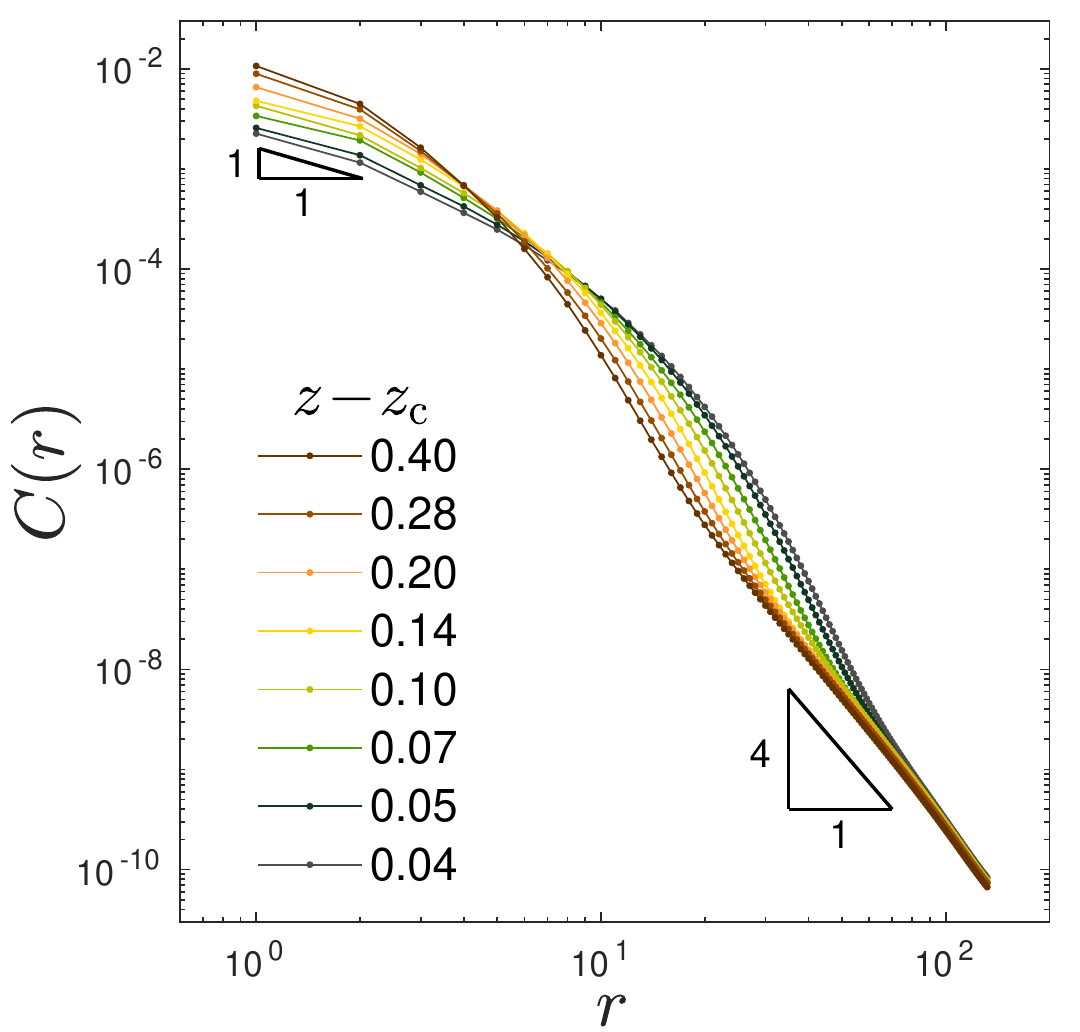}
 \caption{\footnotesize Response functions $C(r)$ are shown to decay as $r^{-1}$ for $r\!\ll\!\xi$ and as $r^{-4}$ for $r\!\gg\!\xi$, as expected.}
 \label{fig:plain_vanilla}
 \end{figure}

For completeness, we present in Fig.~\ref{fig:plain_vanilla} the response function $C(r)\!\sim\! \langle{\bm u}({\bm r})\!\cdot\!{\bm u}({\bm r})\rangle$ (for various level of connectivity $z$), where $\langle\cdot\rangle$ stands for an angular average --- see also the main text ---, as measured in our disordered networks of Hookean springs. $C(r)$ indeed reveals two different power laws at small and large $r$'s, which are more cleanly quantified by analyzing $c(r)$, defined above and presented in Fig.~1b,c in the main text.

\section{C\lowercase{omputer disordered networks}}

We created disordered networks of 16 million nodes each, composed of relaxed Hookean springs connecting (unit) point masses, with both positional and topological (i.e.~degree of connectivity) disorder. This is achieved by adopting the interaction networks of simple, three-dimensional (3D) soft-spheres glasses (see Ref.~\cite{cge_paper_SM} for a description of the soft-spheres model), where we place a Hookean spring between every pair of interacting particles in the original glass.

This procedure results in a disordered spring network of initial coordination $z\!\approx\!16$, which is much larger than the Maxwell threshold $z_{\rm c}\=6$ in 3D. We then systematically remove bonds (springs) by considering in each iteration the bond $i,j$ whose combined connectivity $z_i\!+\!z_j$ is largest. Since there are many bonds that share the same combined connectivity $z_i\!+\!z_j$, we consider a secondary bond-removal criterion: amongst all bonds $i,j$ whose $z_i\!+\!z_j$ is maximal, we select to remove a bond whose difference $|z_i\!-\!z_j|$ is smallest. These two criteria ensure that the connectivity fluctuations of the resulting disordered spring network are small. This procedure is iteratively applied until a target connectivity $z$ is reached. We present the bond dilution algorithm in great detail next.

\subsection{Network dilution algorithm}
\label{sec:dilution}

We assume having in hand a soft-sphere-packing-derived, highly coordinated initial network of $N$ nodes, and a set of edges ${\cal E}$. An edge $e^{ij}\!\in\!{\cal E}$ connects between a pair of neighboring nodes $i,j$, with ranks $z^i$ and $z^j$ respectively. For every edge $e^{ij}$ we define the sum of ranks $s^{ij}\! =\! z^i\! +\! z^j$ and the absolute difference $d^{ij}\! =\! |z^i\! -\! z^j|$.

 \begin{figure}[ht!]
 \includegraphics[width = 0.50\textwidth]{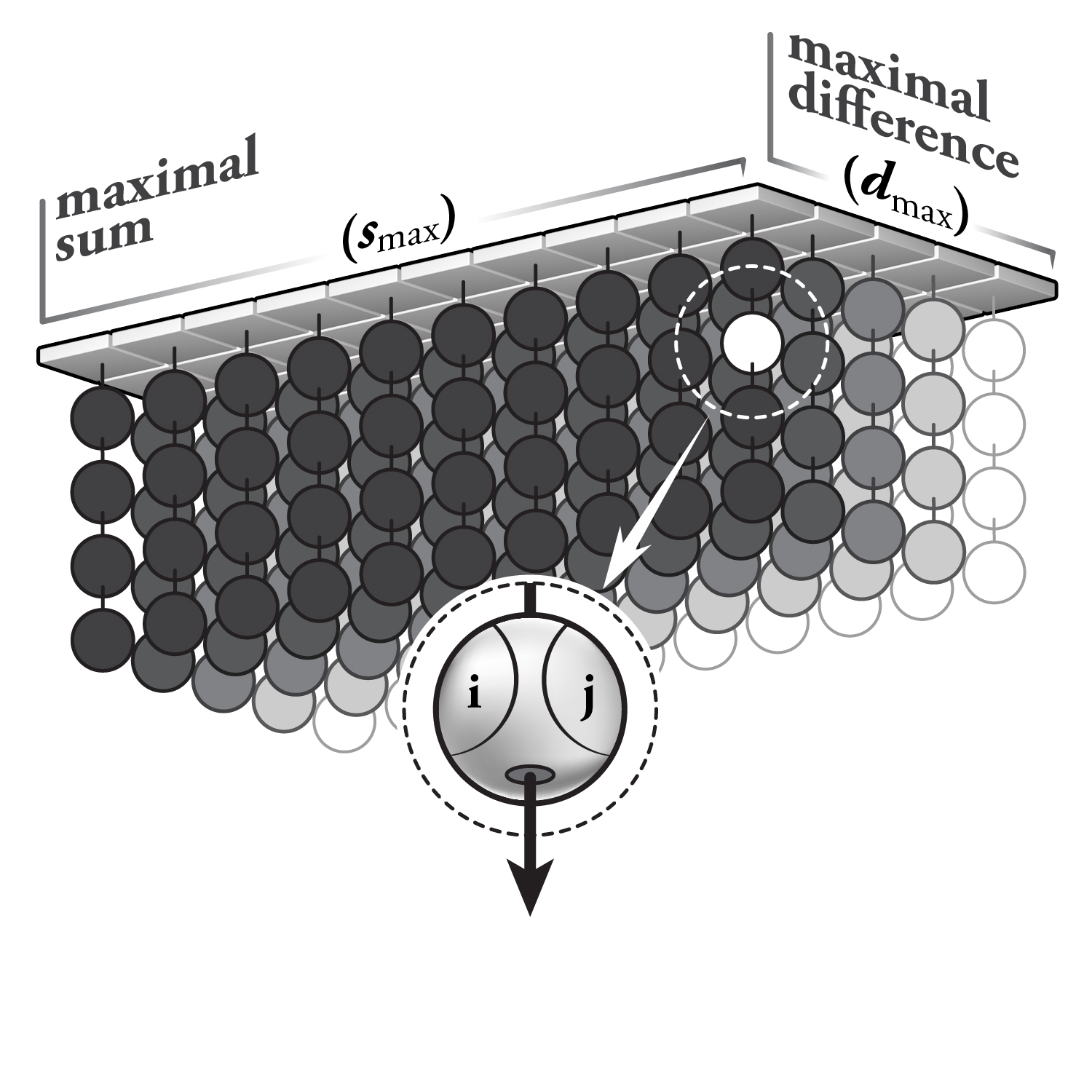}
 \caption{\footnotesize Illustration of the data-structure used by the algorithm for creating homogeneous random networks. Each link holds the indices $i,j$ of an edge $e^{ij}$, in addition to a pointer to the next link. See text for further details.}
 \label{fig:illustration}
 \end{figure}

The algorithm first pre-processes the initial network as follows: each edge in the network is represented by a link which is stored in a data-structure --- illustrated in Fig.~\ref{fig:illustration} --- that is tailored for efficient access to edges with respect to the sum and absolute difference of their ranks during the dilution process. The data-structure consists of a two dimensional array-of-linked-lists ${\cal A}$ of dimensions $s_{\mbox{\scriptsize max}}\!\times \!d_{\mbox{\scriptsize max}}$, where $s_{\mbox{\scriptsize max}}$ is the maximal sum of ranks, and $d_{\mbox{\scriptsize max}}$ is the maximal absolute difference of ranks, amongst all of the network edges. Each list is a concatenation of links that represent edges; each link holds a pointer to the next link in a list, and the two indices $i$ and $j$ that define the edge $e^{ij}$ represented by that link, as illustrated in Fig.~\ref{fig:illustration}. Links that share same sum of ranks $s$ and same absolute difference $d$ are concatenated into a single list which is anchored at ${\cal A}(s,d)$. The pre-processing is described by the following pseudo-code:
\begin{algorithm}[H]
\caption{Pre-processing : $\mathcal O (N)$}
\begin{algorithmic}
\State Assign an $s_{\mbox{\scriptsize max}}\times d_{\mbox{\scriptsize max}}$ array-of-lists ${\cal A}$
\For{each edge $e^{ij} \in {\cal E}$}
  \State $s^{ij} \leftarrow z^i+z^j$
  \State $d^{ij} \leftarrow |z^i-z^j|$
  \State Add chain link $L(i,j)$ at ${\cal A}(s^{ij},d^{ij})$
\EndFor
\end{algorithmic}
\end{algorithm}

We can now proceed with deriving a network with some desired average connectivity $\bar{z}_f$ from the initial network using the production stage of the procedure. The main idea is to proceed as follows: we will start from the first non-empty list ${\cal A}(s,d)$ with the highest index $s$ and lowest index $d$; while a list is found and is non-empty, we will remove the first link from that list, and check for its validity in terms of its attributes $s^{ij}$ and $d^{ij}$ (the two latter could have changed by previous edge removals). If valid, we remove the represented edge from the network, or, otherwise, we will re-insert the link into the appropriate list. In this way edges are consecutively removed from the network until reaching the target mean connectivity $\bar{z}_f$, following the min-max scheme discussed above.

The number of edges which need to be removed from the initial network in order to reach $\bar{z}_f\!<\!\bar{z}$ is $k\!\equiv\!\lfloor N (\bar{z}\! -\! \bar{z}_f)\rfloor$, where $\bar{z}$ is the mean connectivity of the initial network. The production stage is described by the following pseudo-code: \newline

\begin{algorithm}[H]
\caption{Production : $\mathcal O (N)$}
\begin{algorithmic}
\State $s \leftarrow s_{\mbox{\scriptsize max}}$
\State $d \leftarrow 0$
\State {\it counter} $\leftarrow 0$
\While {{\it counter} $< k$}
  \While {${\cal A}(s,d)$ is empty}
    \If {$d<d_{\mbox{\scriptsize max}}$}
        \State Increase $d$
    \Else
        \State Decrease $s$
        \State $d \leftarrow 0$
    \EndIf
  \EndWhile
  \State Read $L(i,j)$ at ${\cal A}(s,d)$
  \If {($z^i + z^j = s$){\ \ \bf AND\ \ }($|z^i-z^j| = d$)}
    \State Remove $L(i,j)$ from ${\cal A}(s,d)$
    \State Update $z^i$ and $z^j$, and remove $e^{ij}$ from the network
    \State Increase {\it counter}
  \Else
    \State Insert $L(i,j)$ to ${\cal A}(z^i+z^j, |z^i-z^j|)$
  \EndIf
\EndWhile
\end{algorithmic}
\end{algorithm}

\vspace{-0.2cm}

To estimate the complexity of this algorithm, we consider the worst case scenario in which each link visits all of the $s_{\mbox{\scriptsize max}}\!\times\! d_{\mbox{\scriptsize max}}$ lists before removal, then the running time would be $\propto\! s_{\mbox{\scriptsize max}}\!\times\! d_{\mbox{\scriptsize max}}\!\times\! N$. However, since both parameters $s_{\mbox{\scriptsize max}}$ and $d_{\mbox{\scriptsize max}}$ are independent of $N$, the complexity remains ${\cal O}(N)$.

%

\end{document}